\documentclass[10pt,twocolumn,a4paper]{article}
\usepackage[T1]{fontenc}
\usepackage[cp1250]{inputenc}
\usepackage{graphicx}
\usepackage{epstopdf}

\usepackage{varioref}% nice with \autoref{label}, use \vpageref[]{label}
\usepackage{amsmath,amssymb,amsfonts}

\usepackage[pdftex,breaklinks=true]{hyperref}
\hypersetup{
colorlinks=true,
linkcolor=black,          % color of internal links
citecolor=black,        % color of links to bibliography
filecolor=black,      % color of file links
urlcolor=black,           % color of external links
bookmarks=true,
bookmarksnumbered=true,
pdfauthor={Piotr Bania},
pdfstartview=FitH,
pdfsubject={Generic Unpacking of Self-modifying, Aggressive, Packed Binary Programs},
pdftitle={Generic Unpacking of Self-modifying, Aggressive, Packed Binary Programs}
}

\usepackage{url}
\usepackage{algorithmic}
\usepackage{listings} 
\usepackage{color} 
\usepackage[ruled,vlined]{algorithm2e}

\title{{Generic Unpacking of Self-modifying, Aggressive, Packed Binary Programs}}
\author{Piotr Bania\\
\texttt{\href{mailto:bania.piotr@gmail.com}{bania.piotr@gmail.com}}}
\date{March 2009}

\begin{document}
\maketitle

\begin{abstract}
Nowadays most of the malware applications are either packed or protected. This techniques are applied especially to evade signature based detectors and also to complicate the job of reverse engineers or security analysts. The time one must spend on unpacking or decrypting malware layers is often very long and in fact remains the most complicated task in the overall process of malware analysis. In this report author proposes MmmBop as a relatively new concept of using dynamic binary instrumentation techniques for unpacking and bypassing detection by self-modifying and highly aggressive packed binary code. MmmBop is able to deal with most of the known and unknown packing algorithms and it is also suitable to successfully bypass most of currently used anti-reversing tricks. This framework does not depend on any other 3rd party software and it is developed entirely in user mode (ring3). MmmBop supports the IA-32 architecture and it is targeted for Microsoft Windows XP, some of the further deliberations will be referring directly to this operating system. 
\end{abstract}

\section{Introduction}
Most of the currently popular malware is runtime packed, encrypted or obfuscated. However not only malware is packed\footnotemark, packers are also successfully used in other popular software applications mostly to defend against cracking and illegal copying. Therefore solutions limited only to packer detection cannot typify whether a packed application is really a malware or not, because such assumption leads to large number of false-positives alerts.
\footnotetext{Author uses the term \emph{packed} and its variations to refer to the techniques of compressing, encrypting (armoring) and obfuscating binary code.}In other words this means that in most of cases detecting, analysing of packed binary code can be only performed after the payload is unpacked. Appending to various external sources  \cite{packed_stats,packed_stats2} about 79\% of malware is packed, where the most popular packers are UPX (more then 50\% of malware files), PECompact, Upack, tElock, Yoda's Crypter, FSG, PESpin, ASPack. Using packing programs causes a transform of original program into a packed program (the original code is compressed, encrypted or both). Each packed program is equipped with so called loader stub (restoration routine) which works before original program. The restoration routine task is to unpack (restore) original packed binary code and throw the execution to original entry point. Each packer typically provide its own loader stub which relies on usage of specific algorithms and because of that it's hard to create one ultimate unpacker which could handle different loader stubs. Furthermore some of the packers like tElock, PESpin, Yoda's Crypt are creating an armored loader stub, which takes an advantage of massive amounts of anti-debugging, anti-reversing tricks and self-modifying code techniques. Such protection techniques often cause a major inconvenience in the malware unpacking and analysis process. \newline
{\noindent}\newline{}This paper will present the method for bypassing packed, obfuscated, armored layers and a couple of methods for finding original entry point (OEP). Author will also try to present unpacking mechanism used in MmmBop, its main goals, limitations and also other related work.

\section{Main Goals}
Like most of the currently known unpackers MmmBop was developed to fulfill specific objectives, which are:
\begin{itemize}
    \item finding original entry point (OEP) and stoping the execution at its place (instrument only the loader stub)
    \item bypassing the protection layers equipped with anti-reversing tricks, obfuscated and self-modifying code, keeping high level of transparency (avoiding interferences)
\end{itemize}
{\noindent}As it was previously stated MmmBop is completely userland application and it does not  interfere with the stability of operating system. It also does not use debugging API, virtual machine or emulation which significantly decrease the risk of being detected. MmmBop uses dynamic binary instrumentation for tracing the execution flow, next section presents its general architecture.

\section{Architecture}
MmmBop consist of two separate modules: Injector and DBI Engine. Each of the modules will be presented in the next subsections. 

\subsection{Injector module}
The main tasks of the Injector module are:
\begin{enumerate}
    \item creating a suspended process of the target application (application to be unpacked) 
    \item loading DBI Engine into the process space
    \item informing DBI Engine about current program entry point
    \item throwing execution to the DBI generated block 
\end{enumerate}

{\noindent}It is important to notice that the entire injection process is done virtually without physical file modification. It especially prominent when the loader stub of the packer is aggressive and computes checksums from the originally packed file. Injector module consist of an own position independent stub, which performs the DBI Engine loading in the target process space. When the Injector work is done it terminates itself and resumes the target process. 

\subsection{DBI Engine}
This module is in fact the heart of MmmBop. It is completely independent and does not rely on any other known dynamic binary instrumentation frameworks like DynamoRIO \cite{dynamorio_page} or Pin \cite{pin_homepage}. Even though those two mentioned DBI frameworks are far more advanced when it comes to instrumenting normal applications (not packed), they were not designed to work with self-modifying, aggressive binary code. Pin authors claim that it supports self-modifying code, unfortunately the tests show that it is still unable to instrument many loader stubs - like the one produced by tElock or PESpin. Furthermore it also contains some other logic errors which often make the instrumentation impossible and because of that it cannot be used in unpacking process directed for aggressive loaders (this will be discussed further in \autoref{prefix_attack}). It appears that Saffron \cite{saffron} (an unpacking approach using Pin) is also unable to work with aggressive packers like tElock. Pin's engine is not open source so it is hard to locate potential errors and address a proper fix. Keeping in mind the DBI limits presented above author managed to create own instrumentation framework, which was developed specifically to instrument the loader stub.\newline

{\noindent}General DBI Engine architecture is presented below (\autoref{img:dbi_arch}):
\begin{figure}[tbhp]
\centering
\includegraphics[scale=0.7]{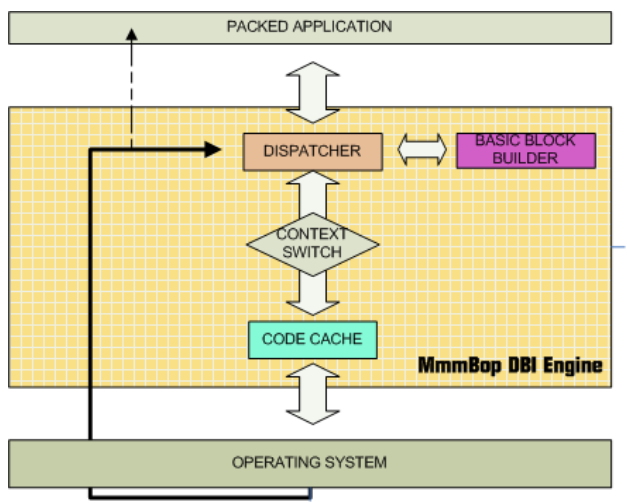}
\caption{General composition of MmmBop Dynamic Binary Instrumentation engine.}
\label{img:dbi_arch}
\end{figure}

{\noindent}In current implementation MmmBop supports only single threaded loader stubs (restoration routines) which is enough to handle most of the known packers.

\subsubsection{Code cache}
Code cache is responsible for enabling native code execution instead of performing emulation. Such solution significantly decreases the slowdown rate (pure emulation is typically about few hundreds times slower then native code execution). Unlike the mechanisms used in DynamoRIO \cite{dynamorio_thesis}, MmmBop code cache stores only one block at a time and also it does not apply any other optimizations like direct/indirect branches linking\footnotemark\ or building traces. This is one of the main assumptions of MmmBop, even if such optimizations considerably increase the speed of instrumented applications they are hard to implement in self-modifying, aggressive code, like packers restoration routines. That's why MmmBop limits such optimizations to minimum and performs them only for indisputable situations (ie. when the basic block is not considered self-modifying). Cached code contains the same logic as the original application code, the only valuable changes are made to the control transfer instructions, which are modified to ensure that MmmBop will always retain control before the code will execute new basic block. Additionally some other instrumentation is injected to the cached code as well, this will be presented more deeply in next sections of this article. 

\subsubsection{Basic Block Builder}
Basic Block Builder like the name says is responsible for creating basic blocks. Basic blocks are a sets of instructions finalized with a single control transfer instruction (in other words basic block contains set of instructions which have a single point of entry and a single point of exit for program control flow). General algorithm for creating a single basic block from original application code is defined as follows (Algorithm \ref{algo_basicblock}).

\footnotetext{With one exception, the links are generated when specified  basic block is considered not self-modifying and the branch target is located in the same basic block.}

\begin{algorithm}
\label{algo_basicblock}
\dontprintsemicolon
\SetKwInOut{Input}{input}
\SetKwInOut{Output}{output}
\Input{$org_{va}$}
\Output{A cached basic block}
\Begin{
$done \longleftarrow false$\;
$current_{va} \longleftarrow org_{va}$\;
\While{$!done$}{\;
$instr$ = disassemble($current_{va}$)\;
\Switch{$instr.type$}{
\uCase{Control Transfer Instruction}{
AddRetainControlInstrumentation()\;
$done \longleftarrow true$\;
}
\Other{
StoreInstruction()\;
break\;
}
}
$current_{va}$ += $instr.len$\;
}
}
\caption{Basic block generation}
\end{algorithm}

{\noindent\newline}{\textbf{Important Note:}}
When generating basic blocks for aggressive binary code special care must be taken because sometimes the input memory address $org_{va}$ or any particular instruction following it maybe invalid. Therefore entire basic block generation process must be protected by exception handler, which will break the process if the source instruction is unavailable and additionally it will not cause a fatal fault in the DBI engine.

\subsubsection{Context switch}
Context switch is essential for separating original program CPU state from MmmBop internal mechanisms. In other words all original registers, flags, stack space are completely separated from the MmmBop. In this case full stack transparency is achieved. There is one interesting (bonus) detail: when working with completely pure stack space (used for switching with the original stack space), it is important to update the {\emph{top of stack}} and {\emph{bottom of stack}} values in the Thread Information Block (TIB) because otherwise the exception handlers (like the one used in basic block builder) will not get executed (and this should be considered as fatal).

\subsubsection{Dispatcher}
As it was previously stated special instrumentation is used for control transfer instructions to make sure MmmBop will retain control before executing new basic block. In fact in such situations the control retains to the Dispatcher element, which decides what to do next. Typically Dispatcher executes Basic Block Builder to create new basic block pointed by the original destination of control transfer instruction. After this is done execution is transfered to the newly generated cached code (of course after performing the context switch).\newline

{\noindent}Besides the presented dispatcher, MmmBop uses few more to follow the original execution process correctly. Those dispatchers will be presented in next sections.

\subsubsection{Exception Dispatcher}
Causing (generating) exceptions is a very popular trick among PE file protectors. Typical scenario works as follows:
\begin{enumerate}
    \item Setup Structured Exception Handler
    \item Generate Exception (execution is thrown to SEH frame)
    \item In SEH frame: check the {\texttt{EXCEPTION\_RECORD}} \cite{msEXCEPTIONRECORD} and {\texttt{CONTEXT}} \cite{msCONTEXT} structures, basing on the values decide what to do next.  
\end{enumerate}
 
{\noindent}When the exception is generated the {\texttt{EXCEPTION\_RECORD}} and {\texttt{CONTEXT}} structures are filled respectively. Whenever the exception will happen in the cached code, the {\texttt{ExceptionAddress}} and {\texttt{Context.Eip}} fields will be filled too, however they will point to the cached code not to the original address. Whenever loader stub makes use of those two values it is almost certain that the instrumentation process will fail. To address this issue MmmBop hooks the {\texttt{KiUserExceptionDispatcher}} \cite{kiuser_blog_entry} function, which is called before the execution of the actual structured exception handler. This gives MmmBop the opportunity to filter and fix the {\texttt{ExceptionAddress}} entry from {\texttt{EXCEPTION\_RECORD}} and {\texttt{Eip}} entry from {\texttt{CONTEXT}} structure. If the exception happened in cached code, MmmBop will calculate the corresponding location and update both of mentioned entries with pointer to original exception address (since both structures are located in writeable memory this is a fairy easy step). Unlike Saffron no kernel module is developed to filter generated exceptions, this solution has it good and bad sides. Special care should be taken with hiding the hook more deeply since loader stub may look for it. However this solution worked perfectly with all tested packers.

\subsubsection{Continue Dispatcher}
Some packers use {\texttt{NtContinue}} \cite{ntcontinue} function to transfer the execution to other location (for example Yoda's Crypter uses this method to return the execution to the original entry point (OEP)). This function is also executed when exception handler returns {\texttt{EXCEPTION\_CONTINUE}} status. Since this function will change the thread context indirectly, MmmBop will loose the track of the execution chain. To resolve this issue MmmBop hooks {\texttt{NtContinue}}, saves old {\texttt{Context.Eip}} and updates it with its own handler. After executing {\texttt{NtContinue}} the control is thrown to MmmBop handler which continues the instrumentation from previously saved {\texttt{Context.Eip}}.  

\section{Unpacking Issues}
In this section author will try to describe some of the most important issues that were necessary to solve to make MmmBop effective. Sometimes to illustrate specific issue more deeply additional real world examples will be provided as well. 

\subsection{Instrumenting CALL}
\label{instrumentingcall}
The {\texttt{CALL}} instruction saves procedure linking information on the stack (return address) and calls defined procedure. Besides the normal usage, self-modifying code uses this instruction to address relatively to the return address placed on the stack, this is often referred as GetPC code (the variant of $\{${\texttt{CALL/POP reg/SUB reg,IMM32}$\}$ instructions is sometimes named as {\emph{delta handling}}). Following code (Listing \ref{listing_delta}) presents how PESpin uses {\texttt{CALL}} instruction for relative addressing. \newline

 {\ttfamily{\footnotesize{
\lstset{language={[x86masm]Assembler}}
\begin{lstlisting}[frame=trbl, label=listing_delta, caption={Fragment of PESpin code, that illustrates using {\texttt{CALL}} return address as a operand for relative addressing and self code modification.}, captionpos=b]{}
004040D8 CALL 2.004040DD
004040DD MOV EBX,DWORD PTR SS:[ESP]
004040E0 ADD EBX,12
004040E3 SUB DWORD PTR DS:[EBX],6B1E8
\end{lstlisting} 
}}} 

{\noindent}{\texttt{CALL}} instruction at {\texttt{0x004040D8}} transfers the execution to {\texttt{0x004040DD}} and simultaneously pushes the return address ({\texttt{0x004040DD}) on the stack. This address is then loaded into {\texttt{EBX}} register (instruction at {\texttt{0x004040DD}}) and increased with {\texttt{0x12}} (instruction at {\texttt{0x004040E0}}). The {\texttt{EBX}} register (now containing {\texttt{0x004040EF}} value) is used via the {\texttt{SUB}} instruction at {\texttt{0x004040E3}} to decode (self-modify) instruction bytes located at {\texttt{0x004040EF}}. Therefore the return address must point to original code location, not the corresponding location in code cache, because even though the cached code does keeps the same program logic it is extended with instrumentation instructions and it is limited to one basic blcok. In other words the same decoding process in reference to cached code may provide other (unstable) results, so executing the instruction located at {\texttt{0x004040E3}} may be fatal in this case. MmmBop takes care of this situation and points the return address to the original location. \newline

{\noindent}While working with self-modifying code it is good to not assume that like in normal application after every {\texttt{CALL}} instruction, {\texttt{RET}} instruction will be used to retain control. In such situations it is quite possible that the execution will never land to the return address stored by {\texttt{CALL}} in fact this is a pretty known trick for disabling the functionality of STEP OVER in debuggers.   

\subsection{Handling self-modifying code}
Since MmmBop only processes one basic block at a time, instruction which modifies memory corresponding to different basic block is simply ignored. However the problems start when an instruction modify memory in range of current basic block. This means that the basic block located in code cache does not correspond to the original one any longer (since it was modified) - and the general logic is probably changed. Most of aggressive protectors make use of this technique like PESpin (see Listing \ref{pespin_self} - extended previous listing).\newline

 {\ttfamily{\footnotesize{
\lstset{language={[x86masm]Assembler}}
\begin{lstlisting}[frame=trbl, label=pespin_self, caption={STAGE1: Fragment of PESpin code, illustrating code before self-modification.}, captionpos=b]{}
004040D7   PUSHAD
004040D8   CALL 2.004040DD
004040DD   MOV EBX,DWORD PTR SS:[ESP]
004040E0   ADD EBX,12
004040E3   SUB DWORD PTR DS:[EBX],6B1E8
004040E9   DEC BYTE PTR DS:[EBX-3]
004040EC   SUB BYTE PTR SS:[ESP],17
004040F0   OUT 46,AL 
004040F2   ADD BYTE PTR DS:[EBX],CL
004040F4   IN AL,74
004040F6   SAHF
004040F7   JNZ SHORT 2.004040FA
\end{lstlisting}
}}} 

{\noindent}As it was previously explained instruction at {\texttt{0x004040E3}} will cause a memory modification pointed by {\texttt{EBX}} register {\texttt{0x004040EF}}. Next instruction located at {\texttt{0x004040E9} will also cause a memory modification to the area {\texttt{0x004040EC}}. This will cause the modification of the basic block logic, now it presents following instructions (Listing \ref{pespin_self_2})\newline. 

 {\ttfamily{\footnotesize{
\lstset{language={[x86masm]Assembler}}
\begin{lstlisting}[frame=trbl, label=pespin_self_2, caption={STAGE2: Fragment of PESpin code, illustrating code after self-modification.}, captionpos=b]{}
004040EC   SUB DWORD PTR SS:[ESP],2.0040342F
004040F3   OR ESP,ESP
004040F5   JE SHORT 2.00404095
004040F7   JNZ SHORT 2.004040FA
\end{lstlisting}
}}} 

{\noindent}Instruction located at {\texttt{0x004040EC}} is completely different then the one before performing decoding process. In addition to \autoref{instrumentingcall}, if the return address placed by {\texttt{CALL}} instruction would be not faked properly, PESpin stub would use the wrong value for further unpacking process (this would result in fault).\newline

{\noindent}To resolve such situations additional instrumentation was used. Typically there are two ways of facing such problems both rely on instrumenting instruction which refers to memory in write mode:
\begin{enumerate}
    \item monitor memory writes and check if the destination memory is located in the range of original basic block
    \item monitor memory writes and check if the original basic block checksum has changed
\end{enumerate}  

{\noindent}Both of the listed mechanisms are implemented in MmmBop and both are comparable in the terms of speed. First mechanism requires some additional code instrumentation since generally the requested memory address can not be statically calculated. After the execution of 'memory write instruction' MmmBop dispatcher checks if it affected current basic block. When MmmBop detects such action, it breaks the current basic block and creates new one (starting after the last executed instruction). Second solution does not require additional instrumentation code for calculating the destination memory address. So after every creation of basic block, a checksum is generated from original code (here the partial Adler-32\footnotemark \cite{adler32} is used as a checksum algorithm). Every time memory write occurs in the basic block, the checksum is calculated one more time from the original basic block code and then it is compared with the previously calculated one. If there is a difference the currently cached basic block is destroyed and next one is generated from the beginning of last instruction that caused the memory write. Current MmmBop implementation enables using one of the two presented methods. The speed comparison between those two mechanisms will be presented in Testimonials section (\autoref{sec:testy}).\newline
\footnotetext{Author is aware of Adler-32 checksum algorithm weaknesses (ie. forging), however they don't represent a important issue in this case.} 

{\noindent}On the side note it's obvious that Prefetch Input Queue (PIQ) \cite{prefetchinputque} tricks like the one used in PESpin ({\texttt{REP STOSB}} instruction used to overwrite itself) have no influence on MmmBop.

\subsection{Prefixes}
\label{prefix_attack}
Special care should be taken while instrumenting control transfer instructions which are encoded together with IA-32 prefixes. Some of the prefixes encoded together with control transfer instruction are used deliberately to cause exceptions (like for example LOCK ({\texttt{0xF0}}) or OPERAND-SIZE ({\texttt{0x66}}) prefix). On the side note Pin tends to ignore such prefixes, such assumption makes it vulnerable to such attacks.

\subsection{Hardware breakpoints}
The IA-32 architecture provides special sets of registers called debug registers, used by the processor for debugging purposes. Those registers allow setting various debug conditions associated with four debug addresses written in {\texttt{DR0-DR3}} registers where the breakpoint condition is stored in the {\texttt{DR7}} register. Unlike software breakpoints, hardware breakpoint (often called as debug breakpoints) do not require changing the original code. The tElock protector makes a pretty nasty usage of this feature, following code illustrates how tElock restoration routine setups hardware breakpoints (Listing \ref{telock_hb_setup}).\newline

{\ttfamily{\footnotesize{
\lstset{language={[x86masm]Assembler}}
\begin{lstlisting}[frame=trbl, label=telock_hb_setup, caption={Fragment of tElock restoration routine, which setups debug breakpoints.}, captionpos=b]{}
00404120   MOV EAX,DWORD PTR DS:[ECX+B4]
00404126   LEA EAX,DWORD PTR DS:[EAX+24]
00404129   MOV DWORD PTR DS:[ECX+4],EAX
0040412C   MOV EAX,DWORD PTR DS:[ECX+B4]
00404132   LEA EAX,DWORD PTR DS:[EAX+1F]
00404135   MOV DWORD PTR DS:[ECX+8],EAX
00404138   MOV EAX,DWORD PTR DS:[ECX+B4]
0040413E   LEA EAX,DWORD PTR DS:[EAX+1A]
00404141   MOV DWORD PTR DS:[ECX+C],EAX
00404144   MOV EAX,DWORD PTR DS:[ECX+B4]
0040414A   LEA EAX,DWORD PTR DS:[EAX+11]
0040414D   MOV DWORD PTR DS:[ECX+10],EAX
00404150   XOR EAX,EAX
00404152   AND DWORD PTR DS:[ECX+14],FFFF0FF0
00404159   MOV DWORD PTR DS:[ECX+18],155
\end{lstlisting}
}}} 

{\noindent}Instructions at {\texttt{0x00404129}}, {\texttt{0x00404135}}, {\texttt{0x00404141}}, {\texttt{0x0040414D}} write the breakpoint location to the {\texttt{DR0}}-{\texttt{DR3}} debug registers respectively. Instruction at  {\texttt{0x00404152}} updates the {\texttt{DR6}} registers and finally instruction at {\texttt{0x00404159}} enables four hardware breakpoints by setting the {\texttt{DR7}} register bits. Since direct changes to debug registers require ring0 privileges, following code executes in the exception handler and it operates on the {\texttt{CONTEXT}} structure ({\texttt{ECX}}). The modified context is then passed to {\texttt{NtContinue}} function which applies it to the selected thread, after resuming the execution selected hardware breakpoints are set. After the CPU will execute instruction corresponding the breakpoint address {\texttt{EXCEPTION\_SINGLE\_STEP}} will be thrown. This exception is filtered by tElock exception handler and following code is executed (see Listing \ref{telock_hb_handle}).\newline
{\ttfamily{\footnotesize{
\lstset{language={[x86masm]Assembler}}
\begin{lstlisting}[frame=trbl, label=telock_hb_handle, caption={Fragment of tElock restoration routine, which handles single step exceptions.}, captionpos=b]{}
00404113   CALL 2.00404119
00404118   DB  00
00404119   POP EAX
0040411A   INC BYTE PTR DS:[EAX]
0040411C   SUB EAX,EAX
0040411E   JMP SHORT 2.00404160
\end{lstlisting}
}}}

{\noindent}Whenever tElock handles single step exception, it increases the byte variable located at {\texttt{0x00404118}} (which is in fact a counter) and it resumes the execution afterwards. This counter value is used in the further parts of the unpacking process. Therefore whenever hardware breakpoints will not be hit the file will not be unpacked correctly. In this case when the cached code is executed instead of the original one it's obvious that the breakpoints will not be detected and the unpacking process will fail. To correctly handle such situations MmmBop monitors the context passed to {\texttt{NtContinue}} function and writes down all the enabled hardware breakpoints locations. Whenever the basic block builder meets the specified breakpoint location MmmBop simply links current instruction to the original breakpoint address. Because of this mechanism breakpoints are correctly handled and MmmBop retains the program control immediately after the exception is thrown.  

% ---------------------------------------------------------------------------
% OEP Finding
% ---------------------------------------------------------------------------
\section{OEP Finding}
\label{sec:oep_finding}
MmmBop may use different approaches directed for finding original entry point of the packed program. Since it is able to instrument all the instructions which cause memory writes, techniques that rely on this approach (detecting the execution of previously written area) may be applied as well. Currently MmmBop focus on control transfer checks, so whenever the control is returned to a basic block located at specified memory range, MmmBop assumes the original entry points was reached. The memory range used in this process generally corresponds to the borders of first section of the packed file. Since most of the packers do not erase such information this solution plays out quite well. From the other hand packers like uPack merge all of the original sections into one\footnotemark, this \footnotetext{On the side note similar mechanism is used in the JollyRoger virus.} requires some manual guessing of the down border of the original executable section. In the more hard situations it seems to be possible to deliver another assumption, like every unpacked program tends to use API functions delivered by the operating system or by additional libraries. Therefore the first\footnotemark control transfer to outside library may be used for further manual analysis (since typically the API call is located just after original entry point), however this should be treated as an alternative technique because it is not always reliable.\footnotetext{Author assumes that the API calls done by the loader stub are ignored.} Additionally some other techniques may be applied to solve the extra cases (ie. the dual-mappings \cite{dualmappings} problem), for example like intercepting the mapping file API\footnotemark\ ({\texttt{MapViewOfFile}} and related functions) or using the technique similar to the one from PolyUnpack \cite{polyunpack}.
\footnotetext{Of course this may require developing a kernel module because native API \cite{native_api,syscall_shellcode} may be used instead.}

% ---------------------------------------------------------------------------
% TESTY
% ---------------------------------------------------------------------------
\section{Testimonials}
\label{sec:testy}
Following section will present sample results obtained in the process of unpacking (original entry point finding) custom executable by MmmBop. In the tests a sample, 8 192 byte application was used. The packers tested were: UPX ver. 3.03w, WinUpackRY ver. 0.39 final, Yoda's Crypter ver. 1.3 (options: CRC check, anti dumping, clear import information, API redirect), tElock ver. 0.98 (options: debugger detection, IAT-redirection), PESpin ver. 1.32 (options: debugger detection, API redirection, antidump protection, code redirection). For each packing tools file was unpacked 10 times (5 times within the usage Adler checksum algorithm, 5 times within the usage of normal instrumentation) and the average value was calculated. The results are written in Table \ref{table:unpack_results} and also illustrated on the chart below (\autoref{img:wykres}). The number of basic block transfers required by specified packer is presented in Table \ref{table:unpack_transfers}.\newline

\begin{table}[tbhp]
\centering
\begin{tabular}{| l | c | c |}
\hline
Packer Name & $U_{tACRC}$ [s] & $U_{tInstr}$ [s]\\ \hline
PESpin & 4.6084 & 4.5795 \\
tElock & 1.2804 & 1.3159 \\
yC & 0.7611 & 0.9472 \\
UPX & 0.0759 & 0.0789 \\
WinUpackRY & 0.5930 & 0.5964 \\
\hline
\end{tabular}
\caption{Time required by MmmBop to unpack a single file in reference to different packers and methods.}
\label{table:unpack_results}
\end{table}

{\noindent}Where:
\begin{itemize}
    \item $U_{tACRC}$ is the time required to unpack a file while using Adler-32 checksum approach
    \item $U_{tInstr}$ is the time required to unpack a file while using normal instrumentation (without the Adler-32 checksum)
\end{itemize}

\begin{figure}[tbhp]
\centering
\includegraphics[scale=0.6]{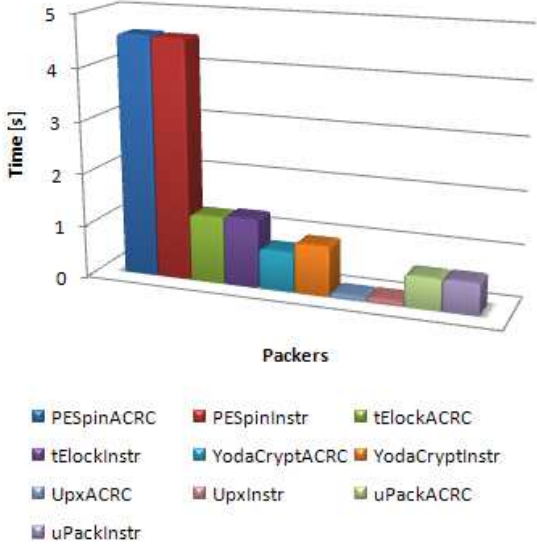}
\caption{Chart illustrating time required by MmmBop to unpack a single file in reference to different packers and methods.}
\label{img:wykres}
\end{figure}

\begin{table}[tbhp]
\centering
\begin{tabular}{| l | c |}
\hline
Packer Name & Basic Block Transfers [\#] \\ \hline
PESpin & 2364189\\
tElock & 311614\\
yC & 180234\\
UPX & 15909\\
WinUpackRY & 131424\\
\hline
\end{tabular}
\caption{Number of basic block transfers in reference to specified packer.}
\label{table:unpack_transfers}
\end{table}

\begin{figure}[tbhp]
\centering
\includegraphics[scale=0.6]{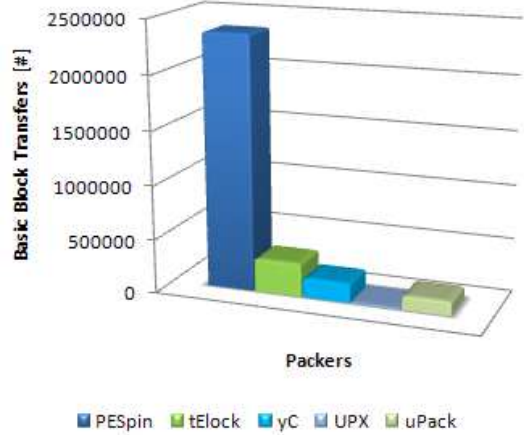}
\caption{Chart illustrating number of basic block transfers in reference to specified packer.}
\label{img:wykres2}
\end{figure}

{\noindent}The results show that both methods used for detecting basic block modification (Adler-32 checksum or the instrumentation approach) produced comparable results. PESpin required the highest amount of time because it uses a lot of control transfers between the basic blocks. It should be obvious that time required for the unpacking process will increase proportionally to the size of packed code (since more iterations will be required to complete the restoration routine). Generally as for the initial MmmBop implementation the results are enough satisfying.  

% ---------------------------------------------------------------------------
% LIMITATIONS
% ---------------------------------------------------------------------------
\section{Limitations}
\label{sec:limitations}
Most of the dynamic binary instrumentation solutions need to modify target process address space. Unfortunately this is unavoidable. Some more sophisticated packers may use this fact for detection purposes. However this may be not so easy to implement, because other software products (like antivirus solutions, firewalls) typically interfere with the target process address space as well (by injecting additional libraries and so on). Packers, especially commercial solutions must work on such machines too, so it is very unlikely such risky solution will be implemented for MmmBop detection. From the other hand packers based on Virtual Machines (VM) approach will not be affected by MmmBop solution. However, in such cases MmmBop may be used for recording the execution trace, which is often very helpful in the further unpacking process.

% ---------------------------------------------------------------------------
% RELATED WORK
% ---------------------------------------------------------------------------
\section{Related Work}
\label{sec:related_work}
There are a number of unpackers available nowadays, this section will try to describe most of the related ones:
\begin{itemize}
    \item OllyBonE \cite{ollybone} is a plugin for OllyDbg \cite{olly_site} which relies on similar mechanism like PaX or Shadow Walker does. It changes the page memory protection of selected region (typically first section) and waits until exception happens at that range. If the exception address (exception EIP) points somewhere inside the protected region then original entry point is found. On the side note similar approach for unpacking purposes was created before by author of this article. The engine was called dEPACKiT and was developed and announced earlier \cite{depackit} - unfortunately it was not released to public. Unlike OllyBonE it was a completely ring3 application. 

     \item Renovo \cite{renovo} uses an emulated environment to
monitor program execution and memory writes. As the emulated environment TEMU is used. Renovo tries to find original entry point by detecting code execution from previously written memory. 

      \item Paradyn Project \cite{paradyn} is a very similar approach to MmmBop. Paradyn uses dynamic binary instrumentation for analyzing packed binary code (it uses Dyninst for this purpose). However it appears to be directed for Unix operating systems and currently it cannot handle self-modifying code. 

      \item Saffron \cite{saffron} also uses dynamic binary instrumentation technique (it uses Pin framework as the dynamic binary instrumentation framework) to monitor program execution together with monitoring memory writes. Additionally it uses hardware paging features in a similar way like OllyBonE and related mechanisms do. Because Saffron relies on Pin framework it is unable to handle such aggressive packers like tElock, PESpin etc.

\end{itemize}

% ---------------------------------------------------------------------------
% FUTURE WORKS
% ---------------------------------------------------------------------------
\section{Future Work}
\label{sec:future_work}
MmmBop is an initial concept of generic unpacker, together with the evolution of the evading and anti-debugging techniques MmmBop must be constantly extended as well. Future MmmBop version should consider handling multi-threading loader stubs and cover more of the aggressive packers. It is quite possible that MmmBop can be quite more optimized the initial version was build without any additional optimizations.

\section{Acknowledges}
Author would like to thank Matt "skape" Miller and Julien Vanegue for helping with writing this article.

% ---------------------------------------------------------------------------
% BILBIOGRAFIA 
% ---------------------------------------------------------------------------

%\newpage
\bibliographystyle{plain}
\bibliography{bibliografia}

\end{document}